\RequirePackage{fix-cm}
\documentclass[smallextended]{svjour3}       
\smartqed  
\usepackage{graphicx}
\journalname{ }
\begin{document}

\title{Two families of  Entanglement-assisted Quantum MDS Codes from cyclic Codes
}


\author{Liangdong Lu        \and
        Wenping Ma  \and
        Ruihu Li \and
        Hao Cao 
}


\institute{ Liangdong Lu \at
              Department of Basic Science,  Air Force Engineering
              University,Xi'an, Shaanxi 710051, China,\\
              School of Telecommunications Engineering,  Xidian University,
               Xi'an, Shaanxi 710051, China,\\
              \email{kelinglv@163.com}           
           \and
            Wenping Ma \at
              School of Telecommunications Engineering,  Xidian University, Xi'an, Shaanxi 710051, China,
           \and
            Ruihu Li \at
              Department of Basic Science,  Air Force Engineering
              University,Xi'an, Shaanxi 710051, China,
               \and
               Hao Cao \at
              School of Information and Network Engineering,
              Anhui Science and Technology University, Chuzhou, 233100, China\\
              School of Mathematical Science, Huaibei Normal University, Huaibei, 235000, China\\
             \email{13655505689@163.com}
           }

\date{Received: date / Accepted: date}

\maketitle

\begin{abstract}
With entanglement-assisted (EA) formalism, arbitrary classical
linear codes are allowed to transform into EAQECCs by using
pre-shared entanglement between the sender and the receiver.
 In  this paper, based on
classical cyclic MDS codes by exploiting pre-shared maximally
entangled states,  we construct  two families  of $q$-ary
entanglement-assisted quantum MDS codes
$[[\frac{q^{2}+1}{a},\frac{q^{2}+1}{a}-2(d-1)+c,d;c]]$, where q is a
prime power  in the form of  $am+l$, and $a=(l^2+1)$ or
$a=\frac{(l^2+1)}{5}$.
 We show that all of $q$-ary EAQMDS have minimum
distance upper limit much larger than the known quantum MDS (QMDS)
codes of the same length. Most of these $q$-ary EAQMDS codes are new
in the sense that their parameters are not covered  by the codes
available in the literature.

\end{abstract}

\section{Introduction}
\label{intro}

Entanglement-assisted quantum error correcting codes (EAQECCs) play an important role in quantum
information processing and quantum computation\cite{shor,cal1}.
Arbitrary classical linear
codes are allowed to transform into EAQECCs by using pre-shared entanglement between
the sender and the receiver under the entanglement-assisted (EA) formalism\cite{bru06}.

Let $q$ be a prime power. A $q$-ary  $[[n,k,d;c]]$ EAQECC  that
encodes $k$ information qubits into $n$ channel qubits with the help
of  $c$ pairs of maximally-entangled Bell states (ebits) can correct
up to $\lfloor\frac{d-1}{2}\rfloor$  errors, where  $d$ is the
minimum distance of the code. A
$q$-ary $[[n,k,d;c]]$ EAQECC is denoted by $[[n,k,d;c]]_{q}$.
Currently, many works have
focused on the construction of EAQECCs based on classical
linear codes, see
\cite{Wil1,Hsi,lai3,lai,lai2,Fujiwara,Hsieh,Wilde,Lu1,guo}.
As in classical coding theory, one of the central tasks in quantum
coding theory is to construct quantum codes and EA-quantum
codes with the best possible minimum distance.

{\bf  Theorem 1.1 \cite{bru06,lai}.} ( EA-Quantum Singleton Bound) An
$[[n,k,d;c]]_{q}$ EAQECC satisfies $$2(d-1)\leq n-k+c,$$ where
$0\leq c \leq n-1$.

 An EAQECC achieving this bound is  a EA-quantum
 maximum-distance-separable (EAQMDS) code.
 According to \cite{Ketkar}, there are no nontrivial MDS stabilizer codes
of lengths exceeding $q^{2}+1$ except when $q$ is even and $d=4$ or
$d=q^{2}$ in which case $n\leq q^{2}+2$.
 Furthermore, it is a very difficult task to construct a quantum MDS code of length $n\leq q^2+1$ with minimal
distance larger  than $q+1$\cite{Jin,Chen,Kai2,ZhangT1}.
Therefore, in order to achieve larger minimal distance,
 one need to construct a EA-quantum MDS code.
The following Proposition is one of the most frequently used
construction methods.

 {\bf Proposition 1.2} \cite{bru06,Wil1}.\ \   If  $\mathcal {C}$$=[n,k,d]_{q^{2}}$
is a classical code over $F_{q^{2}}$ and $H$ is its parity check
matrix, then $\mathcal{C}$$^{\perp _{h}}$ EA stabilizes an $[[n,2k-n+c,d;c]]_{q}$ EAQECC,
 where $c=$rank$(HH^{\dagger})$ is the number of maximally entangled states
required and $H^{\dagger}$ is the conjugate
 matrix of $H$  over $F_{q^{2}}$.

In resent years,  scholars have constructed several
entanglement-assisted quantum codes with good parameters in
\cite{bru06,Wil1,Hsi,lai3,lai,lai2,Fujiwara,Hsieh,Wilde,Qian1,Qian2,Chen2,Liu,Luo1,Galindo,Qian11,tian,pang,Zhu22,Guenda1}.
Many classes of EAQMDS codes have been constructed by  different
methods, in particular, by the Hermitian constructions from cyclic
codes, constacyclic codes or negacyclic
codes\cite{Chen2,Chen3,Liu,Lu22,Lu3}.
 In\cite{Lu1,Li1}, we proposed
 the concept about a decomposition of the defining set of cyclic codes,
 and  construct some good entanglement-assisted
quantum codes with the help of this concept\cite{Li1}. In this
paper, we construct two families of EA-quantum MDS codes with length
$n$ from cyclic codes.
 More precisely, Our main contribution on new
$q$-ary quantum MDS codes is as follows:

(1)  $$[[\frac{q^{2}+1}{a},\frac{q^{2}+1}{a}-2d+6,d;4]]$$

 where $q=am+l$, $a=(l^2+1)$, $l$ is an odd number, and $(l+1)m+3\leq d \leq (3l-4)m+3$ is odd;

(2)  $$[[\frac{q^{2}+1}{a},\frac{q^{2}+1}{a}-2d+6,d;4]]$$

 where $q=am+l$,  $a=\frac{(l^2+1)}{5}$,  $l=10t+3$ or
$l=10t+7$ is an odd number, and
$(\frac{l-1}{2}+\lceil\frac{l}{10}\rceil)m+5\leq d \leq (l+1)m+5$ is
 odd.

In construction (1), consuming four pairs of maximally entangled
states, we obtain a family of EA-quantum MDS codes with the minimal
distance twice larger than the standard quantum MDS codes in
Ref.\cite{Chen,ZhangT1}. Comparing the parameters with all known
EA-quantum MDS codes, we find that these quantum MDS  codes are new
in the sense that their parameters are not covered by the codes
available in the literature.

The paper is organized as follows. In Section 2, basic notations and
results about EA-quantum codes and cyclic codes are provided.
In Section 3, we give new classes of EA-quantum MDS
codes. The conclusion is given in Section 4.

\section{Preliminaries}
\label{sec:1} In this section, we review some basic results on cyclic codes,
BCH codes, and EAQECCs for the purpose of this paper. For details on BCH codes and cyclic
codes can be found in standard textbook on coding theory \cite{Macwilliams,Huffman}, and
 for EAQECCs please see Refs.\cite{bru06,Wil1,Hsi,lai3,lai,Fujiwara,Lu1}.

Let $q$ be a prime power. $F_{q^{2}}$ denotes the finite field with $q^{2}$
elements. For any $\alpha \in F_{q^{2}}$, the conjugation of
$\alpha$ is denoted by $\overline{\alpha}=\alpha^{q}$. Given two
vectors $\mathbf{x}=(x_{1},x_{2},\cdots,x_{n})$ and
$\mathbf{y}=(y_{1},y_{2},\cdots,y_{n})\in F_{q^{2}}^{n}$, their
Hermitian inner product is defined as
$(\mathbf{x},\mathbf{y})_{h}=\sum \overline{x_{i}}y_{i}=\overline{x_{1}}y_{1}+\overline{x_{2}}y_{2}+\cdots+\overline{x_{n}}y_{n}.$
For a linear code $\mathcal{C}$ over $F_{q^{2}}$ of length $n$, the
Hermitian dual code $\mathcal{C}^{\bot _{h}}$ is defined as
 $\mathcal{C}^{\bot _{h}}=\{x\in  F_{q^{2}}^{n} | (x,y)_{h}=0, \forall  y $$\in \mathcal{C}\}$.
If $\mathcal{C}^{\bot _{h}}\subseteq\mathcal{C} $, then
$\mathcal{C}$ is called a Hermitian dual containing code, and
$\mathcal{C}^{\bot _{h}}$ is called a Hermitian self-orthogonal
code.

We now recall some results about cyclic codes. For a cyclic code $\mathcal{C}$, each codeword $c =
(c_{0}, c_{1}, \cdots, c_{n-1})$ is customarily represented in its
polynomial form: $c(x) = c_{0} + c_{1}x + \cdots + c_{n-1}x_{n-1},$
and the code $\mathcal{C}$ is in turn identified with the set of all
polynomial representations of its codewords. The proper context for
studying  cyclic codes is the residue class ring
$\mathcal{R}_{n}$$=\mathbf{F}_{q}[x]/(x^{n}-1)$. $xc(x)$ corresponds
to a cyclic shift of $c(x)$ in the ring $\mathcal{R}_{n}$. As we
all know, a linear code $\mathcal{C}$ of length $n$ over $F_{q^{2}}$
is cyclic if and only if C is an ideal of the quotient ring
$\mathcal{R}_{n}=\mathbf{F}_{q}[x]/(x^{n}-1)$. It follows that
$\mathcal{C}$ is generated by monic factors of $(x^{n}-1)$, i.e.,
$\mathcal{C}=\langle f(x) \rangle$ and $f(x)|(x^{n}-1)$. The $f(x)$
is called the generator polynomial of $\mathcal{C}_{n}$.

 Let $\gamma$ be a primitive $n$-th root of unity in some
splitting field  of $x^{n}-1$ and $T=C_{b}\cup C_{b+1}\cup \cdots
\cup C_{b+\delta-2}$. A cyclic code  $\mathcal{C}$ of length $n$
with generator polynomial $g(x)=\Pi_{i\in T}(x-\gamma^{i})$ is
called a BCH code with designed distance $\delta$, and $T$ is called
the defining set of  $\mathcal{C}$.
Let $s$ be an integer with $0\leq s < n$, the
$q^{2}$-cyclotomic coset modulo $n$ that contains $s$ is defined by
the set $C_{s}=\{s, sq^{2}, sq^{2\cdot 2}, \cdots, sq^{2(k-1)} \}$
(mod $n$), where $k$ is the smallest positive integer such that
$xq^{2k}$ $\equiv x$ (mod $n$).
We can see that the defining set $T$ is a union of some
$q^{2}$-cyclotomic cosets module $n$ and $dim(\mathcal{C}) =
n-|T|$.

Let $\mathcal {C}$ be a cyclic code with a defining set $T = \bigcup
\limits_{s \in S} C_{s}$. Denoting $T^{-q}=\{n-qs | s\in T \}$, then
we can deduce that the  defining set of $\mathcal {C}$$^{\bot _{h}}$
is $T^{\perp _{h}} =$$ \mathbf{Z}_{n}
$$\backslash T^{-q}$, see Ref. \cite{Lu1}.

 A cyclotomic coset $C_{s}$ is {\it skew symmetric } if $n-qs$ mod $n\in
 C_{s}$; and otherwise is skew asymmetric otherwise. {\it  Skew asymmetric
 cosets}
$C_{s}$ and $C_{n-qs}$ come in pair, we use $(C_{s},C_{n-qs})$ to
denote such a pair.

The following results on $q^{2}$-cyclotomic  cosets, dual containing
cyclic codes are bases of our discussion.

According to \cite{Lu1,Li2}, $\mathcal{C}^{\perp _{h}}\subseteq$
$\mathcal{C}$ can be described by the relationship of its cyclotomic
coset $C_{s}$.  In order to construct EA-quantum MDS codes for
larger minimal distance of code length $n\leq q^2+1$, we introduce a
fundamental definition of decomposition of the defining set of
cyclic codes.

{\bf Definition 2.1\cite{Lu1}}  {\it  Let $ \mathcal {C}$ be a cyclic
code of length $n$ with defining set $T$. Denote $T_{ss}=T
\cap$$T^{-q}$ and $T_{sas}=T \setminus
$$T_{ss}$, where $T^{-q}=\{n-qx | x\in T \}$. $T=T_{ss} \cup
T_{sas}$ is called decomposition of the defining set of
$\mathcal{C}$.}

To  determine $T_{ss}$ and $T_{sas}$, we give the following lemma to
characterize them.

{\bf Lemma 2.2 \cite{Li2}.} Let $gcd(q, n) = 1$,
$ord_{rn}$$(q^{2})=m$, $0 \leq x, y$, $z \leq n-1$.

(1) $C_{x}$ is skew symmetric if and only if there is a $t\leq
\lfloor\frac{m}{2}\rfloor$
 such that $x \equiv xq^{2t+1}$(mod n).

(2) If $C_{y}\neq C_{z}$, $(C_{y}, C_{z})$ form a skew asymmetric
pair if and only if there is a $t\leq \lfloor\frac{m}{2}\rfloor$
such that $y \equiv zq^{2t+1}$ (mod n) or $z \equiv yq^{2t+1}$(mod
n).

Using the decomposition of a  defining set  $T$, one can calculate
the number of needed ebits with a algebra method.

 {\bf Lemma 2.3. \cite{Lu1}} Let $T$ be a defining set of a cyclic
code $ \mathcal {C}$, $T=T_{ss}\cup T_{sas}$ be decomposition of
$T$. Using $\mathcal{C}$$^{\perp_{h}}$ as EA stabilizer, the optimal
number of needed  ebits is $c=\mid T_{ss} \mid$.

{\bf Lemma 2.4.} (The BCH bound) Let $\mathcal{C}$ be a  cyclic code
of length $n$ with defining set $T$.
 Assume $T$ contains $d-1$ consecutive elements for some integer $d$.
 Then the minimum distance of $\mathcal{C}$ is at least $d$.

{\bf Theorem 2.5.} Let $\mathcal{C}$ be an $[n,k,d]_{q^{2}}$ MDS
code with defining set $T$,  and the  decomposition of $T$ be
$T=T_{ss}\cup T_{sas}$. Then $\mathcal{C}$$^{\perp_{h}}$  EA
stabilizes an $q$-ary $[[n,n-2|T|+|T_{ss}|,d \geq \delta ;
|T_{ss}|]]$ EA-quantum MDS Code.

\section{New  EA-quantum MDS Codes of Lenght $n=\frac{q^{2}+1}{a}$}

\label{sec:1} In this section, we consider cyclic codes over
$F_{q^{2}}$ of length $n=\frac{q^{2}+1}{a}$ to construct EA-quantum
codes, where $q=am+l$, $a=(l^2+1)$ and $l$ is an odd number. To do
this, we give  a decomposition of the defining set of cyclic codes
over $F_{q^{2}}$ of length $n$. Let $n=\frac{q^{2}+1}{a}$,  and
$s=\frac{n+1}{2}$, where $q=am+l$, $a=(l^2+1)$ and $l$ is an odd
number. Obviously, the $q^{2}$-cyclotomic cosets modulo $n$ are
$$C_{s}=\{s,s-1\},C_{s+1}=\{s+1,s-2\},\cdots,C_{n-2}=\{n-2,2\},C_{n-1}=\{n-1,1\}.$$

\subsection{$q=am+l$, $a=(l^2+1)$ }

\label{sec:2} In this subsection, we assume that $q$ is an odd prime
power of the form $q=am+l$, $a=(l^2+1)$, $l$ is an odd number and
$s=\frac{n+1}{2}$. We construct new $q$-ary EA-quantum MDS codes of
length $n=\frac{q^{2}+1}{a}$ from cyclic codes.

Let us first give a useful lemma for our constructions.

 {\bf  Lemma 3.1:} Let $q=am+l$, $a=l^2+1$, $l$ is an odd number,
 $n=\frac{q^{2}+1}{a}$ and $s=\frac{n+1}{2}$. If $\mathcal{C}$ is a $q^{2}$-ary cyclic code of
length $n$ with define set $T=\bigcup_{i=0}^{k}C_{s+i}$, where
$0\leq k\leq (\frac{3l-1}{2})m$, and the decomposition of a
defining set
 $T=T_{ss}\bigcup T_{sas}$, then
 $T_{ss}=\{C_{s+\frac{l+1}{2}m},C_{s+\frac{l-1}{2}m}\}$, and $|T_{ss}|=4$.

 {\bf  Proof:} Since $-(s+\frac{l+1}{2}m)q\equiv
 -(s+\frac{l+1}{2}m)\cdot(am+l)$ $\equiv \frac{l^2+1}{2}m^2+(l+\frac{l-1}{2})m+1$ $\equiv s+\frac{l-1}{2}m$ (mod $n$),
  $\{C_{s+\frac{l+1}{2}m},C_{s+\frac{l-1}{2}m}\}$ forms a  skew asymmetric pair.

Let $T=\bigcup_{i=0}^{k}C_{s+i}$, where $0\leq k\leq
(\frac{3l-1}{2})m$.
 According to the concept about a  decomposition of the  defining set $T$,
 one obtain that $T_{sas}=T\backslash T_{ss}$. In order to testify $|T_{ss}| =4$ if
$0\leq k\leq (\frac{3l-1}{2})m$, from Definition 2.1 and Lemma 2.2,
 we need  to testify that there is no skew symmetric cyclotomic
 coset, and any two cyclotomic
 coset do not form a  skew asymmetric pair in  $T_{sas}$.
For $x\in \{0,1,2,\cdots,n-1\}$,
 the $q^{2}$-cyclotomic cosets modulo $n$
are  $C_{x}=\{x,n-x\}$. Let $I=\{s+i|0\leq i\leq
(\frac{3l-1}{2})m\}$. We only need  to testy that for $\forall x\in
I$, $-qx$ (mod $n$)$\not \in I$ and
 $T_{ss}=\{C_{s+\frac{l+1}{2}m},C_{s+\frac{l-1}{2}m}\}$. That implies that if $x,y\in I$, from  Lemma
2.2, $C_{x}$ is not a skew symmetric cyclotomic  coset, and any
$C_{x},C_{y}$ do not form a skew asymmetric pair if and only if
$x+yq\not\equiv0$ mod $n$.

Divide $I$ into $(\frac{3l-1}{2})m$ parts $I_{1}=[s,s+m]$,
$I_{2}=[s+m+1,s+2m-1]$, $I_{3}=[s+2m+1,s+3m]$, $\cdots$, and
$I_{(\frac{3l-1}{2})m}=[s+(\frac{3l-1}{2}-1)m+1,s+(\frac{3l-1}{2})m]$.
Since $q=(l^2+1)m+l$, $n=\frac{q^2+1}{l^2+1}=(l^2+1)m^2+2lm+1$ and
$s=\frac{n+1}{2}=\frac{l^2+1}{2}m^2+lm+1$, if $ x,y\in I_{1}$, then
$(\frac{l^2+1}{2}m+\frac{l+1}{2})\cdot
n<(\frac{l^2+1}{2}m+\frac{l+1}{2})n+\frac{l^2+1}{2}m+\frac{l+1}{2}=s(q+1)\leq
x+yq \leq (s+m)(q+1)<(\frac{l^2+1}{2}m+\frac{l+1}{2}+1)\cdot n$;
 if $ x,y\in I_{2}$, then
$(\frac{l^2+1}{2}m+\frac{l+1}{2}+1)\cdot n<(s+m+1)\cdot (q+1)\leq
x+yq \leq (s+2m-1)(q+1)<(\frac{l^2+1}{2}m+\frac{l+1}{2}+2)\cdot n$;
and using this method, one can obtain that
 if $ x,y\in I_{im}$, then
$(\frac{l^2+1}{2}m+\frac{l+1}{2}+i)\cdot n<(s+im+1)\cdot (q+1)\leq
x+yq \leq (s+(i+1)m)(q+1)<(\frac{l^2+1}{2}m+\frac{l+1}{2}+i+1)\cdot
n$, where $0\leq i\leq \frac{3l-1}{2}-1$.

Hence, there is no skew symmetric cyclotomic cosets, and any two
cyclotomic  coset do not form a skew asymmetric pair in $T\setminus
\{C_{s+2m},C_{s+m}\}$. That implies that
$T_{ss}=\{C_{s+\frac{l+1}{2}m},C_{s+\frac{l-1}{2}m}\}$ and
$|T_{ss}|=4$ for  $0\leq k\leq (\frac{3l-1}{2})m$, when the defining
set $T=\bigcup_{i=0}^{k}C_{s+i}$, where $0\leq k\leq
(\frac{3l-1}{2})m$.\\

{\bf  Theorem 3.2:}   Let  $q$ is an odd prime power of the form
$q=am+l$, $a=l^2+1$, $l$ is an odd number, $n=\frac{q^{2}+1}{a}$.
Then there exists a q-ary
$[[\frac{q^{2}+1}{a},\frac{q^{2}+1}{a}-2d+6,d;4]]$- EA-quantum MDS
codes, where $(l+1)m+3\leq d \leq (3l-4)m+3$ is odd.

 {\bf  Proof:} Consider the cyclic codes over $F_{q^{2}}$ of length
$n=\frac{q^2+1}{a}$ with defining set $T=\bigcup_{i=0}^{k}C_{s+i}$,
where $0\leq k\leq (\frac{3l-1}{2})m$, and $q$ is an odd prime power
with the form of $q=am+l$, $a=(l^2+1)$. By Lemma 3.1, there is
$c=|T_{ss}|=4$ if $\frac{l+1}{2}m\leq k\leq (\frac{3l-1}{2})m$.
Since every $q^{2}$-cyclotomic coset $C_{s+x}=\{s+x,s-x-1\}$, $0\leq
x\leq s-1$ and $s=\frac{n+1}{2}$, we can obtain that $T$ consists of
$2(k+1)$ integers $\{s-k-1,\cdots,s-2,s-1,s,s+1,\cdots,s+k\}$. It
implies that $\mathcal{C}$ has minimum distance at least $2k+1$.
Hence, $\mathcal{C}$ is a $q^{2}$-ary cyclic code with parameters
$[n,n-2(k+1),\geq 2k+3]$. Combining Theorem 2.5 with EA-quantum
Singleton bound, we can obtain a EA-quantum MDS code with parameters
$[[\frac{q^{2}+1}{a},\frac{q^{2}+1}{a}-2d+6,d;4]]_{q}$, where
$(l+1)m+3\leq d \leq (3l-4)m+3$ is odd.

\begin{center}
Table 1 EAQMDS codes with $n=\frac{q^{2}+1}{a}$, $a=l^2+1$ \\
\begin{tabular}{lllllllllll}
  \hline
          &q           &$[[n,k,d;4]]_{q}$                  &d     is odd                           \\
\hline
 $q=am+l$, $a=l^2+1$          &   13                   &$[[17,23-2d,d;4]]_{13}$              &$7\leq d \leq 11$                   \\
 \quad\quad $l=3$
                      &    23                   &$[[53,59-2d,d;4]]_{23}$               &$11\leq d \leq 19$               \\

                      &   43                     &$[[185,191-2d,d;4]]_{43}$                &$19\leq d \leq 35$             \\

  \quad\quad$l=5 $                  &   31                   &$[[37,43-2d,d;4]]_{37}$              &$9\leq d \leq 17$              \\

                      &    83                   &$[[265,271-2d,d;4]]_{83}$               &$21\leq d \leq
                      45$\\
                                  &   109                    &$[[457,463-2d,d;4]]_{109}$                &$27\leq d \leq 59$             \\

\quad\quad $l=7$              &   107                   &$[[229,235-2d,d;4]]_{107}$              &$19\leq d \leq 43$              \\

                  &    157                   &$[[493,499-2d,d;4]]_{157}$               &$27\leq d \leq
                  63$\\
            &   257                    &$[[1321,1327-2d,d;4]]_{257}$                &$43\leq d \leq 103$             \\
 \quad\quad $l=9$          &   173                   &$[[365,371-2d,d;4]]_{173}$              &$23\leq d \leq 55$              \\

            &    337                   &$[[1385,1391-2d,d;4]]_{337}$               &$43\leq d \leq 107$\\

            &    419                   &$[[2141,2147-2d,d;4]]_{419}$               &$53\leq d \leq 133$               \\

 \hline
  \end{tabular}
\end{center}

\subsection{$q=am+l$,  $a=\frac{l^2+1}{5}$}

\label{sec:3} In this subsection, we assume that $q$ is an odd prime
power of the form $q=am+l$, $a=\frac{l^2+1}{5}$, $l=10t+3$ or
$l=10t+7$ is an odd number and $s=\frac{n+1}{2}$. We construct new
$q$-ary EA-quantum MDS codes of length $n=\frac{q^{2}+1}{a}$ from
cyclic codes. Obviously, for the cyclic codes, the
$q^{2}-$cyclotomic coset $C_{i}$ modulo $n$ are
$C_{s}=\{s,s-1\},C_{s+1}=\{s+1,s-2\},\cdots,C_{n-2}=\{n-2,2\},C_{n-1}=\{n-1,1\}$.

First, we give a useful lemmas for our constructions.

 {\bf  Lemma 3.3:} Let $q=am+l$, $a=\frac{l^2+1}{5}$, $l=10t+3$ or
$l=10t+7$ is an odd number, $n=\frac{q^{2}+1}{a}$ and
$s=\frac{n+1}{2}$. If $\mathcal{C}$ is a $q^{2}$-ary cyclic code of
length $n$ with define set $T$.

(a) If $l=10t+3$,  $T=\bigcup_{i=0}^{k}C_{s+i}$, where $0\leq k\leq
\frac{l+1}{2}m+1$, and the decomposition of a  defining set
 $T=T_{ss}\bigcup T_{sas}$, then
 $T_{ss}=\{C_{s+\frac{l+3}{4}m},C_{s+\lfloor \frac{l}{10} \rfloor m}\}$, and $|T_{ss}|=4$.

(b) If $l=10t+7$, $T=\bigcup_{i=0}^{k}C_{s+i}$, where $0\leq k\leq
\frac{l+1}{2}m+2$, and the decomposition of a  defining set
 $T=T_{ss}\bigcup T_{sas}$, then
 $T_{ss}=\{C_{s+\frac{(\frac{l-1}{2}+\lceil\frac{l}{10}\rceil)}{2}m},C_{s-\lceil\frac{l}{10}\rceil m-1}\}$,
 and $|T_{ss}|=4$.

 {\bf  Proof:} For $l=10t+3$, since $-(s+\frac{l+3}{4}m)q\equiv
 -(\frac{l^2+1}{10}m^2+lm+2\frac{l^2+1}{5})(\frac{l^2+1}{5}m+l)$  $\equiv s+\lfloor \frac{l}{10} \rfloor m$ mod $n$,
 $\{C_{s+\frac{l+3}{4}m},C_{s+\lfloor \frac{l}{10} \rfloor m}\}$ forms a  skew asymmetric pair.
For $l=10t+7$, since
$-(s+\frac{(\frac{l-1}{2}+\lceil\frac{l}{10}\rceil)}{2}m)q\equiv
 -(\frac{l^2+1}{10}m^2+lm+3+\frac{(\frac{l-1}{2}+\lceil\frac{l}{10}\rceil)}{2}m)(\frac{l^2+1}{5}m+l)$
 $\equiv s-\lceil\frac{l}{10}\rceil m-1$ mod $n$,
 $\{C_{s+\frac{(\frac{l-1}{2}+\lceil\frac{l}{10}\rceil)}{2}m},C_{s-\lceil\frac{l}{10}\rceil m-1}\}$ forms a  skew asymmetric pair.

Let $T=\bigcup_{i=0}^{k}C_{s+i}$, where $0\leq k\leq
\frac{l+1}{2}m+1$ if $l=10t+3$; $0\leq k\leq \frac{l+1}{2}m+2$ if
$l=10t+7$.
 According to the concept about a  decomposition of the  defining set $T$,
 one obtain that $T_{sas}=T\backslash T_{ss}$.
For $x\in \{0,1,2,\cdots,n-1\}$,
 the $q^{2}$-cyclotomic cosets modulo $n$
are  $C_{x}=\{x,n-x\}$. Let $I=\{s+i|0\leq i\leq 7m\}$. We only need
to testy that for $\forall x\in I$, $-qx$ (mod $n$)$\not \in I$ and
$T_{ss}=\{C_{s+\frac{l+3}{4}m},C_{s+\lfloor \frac{l}{10} \rfloor
m}\}$ if  $l=10t+3$;
$T_{ss}=\{C_{s+\frac{(\frac{l-1}{2}+\lceil\frac{l}{10}\rceil)}{2}m},C_{s-\lceil\frac{l}{10}\rceil
m-1}\}$ if  $l=10t+7$.

That implies that if $x,y\in I$, from  Lemma 2.2, $C_{x}$ is not a
skew symmetric cyclotomic  coset, and any $C_{x},C_{y}$ do not form
a skew asymmetric pair if and only if $x+yq\not\equiv0$ mod $n$.

Divide $I$ into $\frac{l+1}{2}$ parts such as $I_{1}=[s,s+m]$,
$I_{2}=[s+m+1,s+2m]$, $\cdots$,
$I_{\frac{l+1}{2}}=[s+(\frac{l+1}{2}-1)m+3,s+\frac{l+1}{2}m+2]$ for
$l=10t+3$; and $I_{1}=[s,s+m]$, $I_{2}=[s+m+1,s+2m]$, $\cdots$,
$I_{\frac{l+1}{2}}=[s+(\frac{l+1}{2}-1)m+2,s+\frac{l+1}{2}m+1]$ for
$l=10t+7$. Since $q=26m+5$, $n=\frac{q^2+1}{26}=10m^2+14m+5$ and
$s=\frac{n+1}{2}=5m^2+7m+3$, if $ x,y\in I_{1}$, then
$(\frac{l^2+1}{2}m+\frac{l+1}{2})\cdot
n<(\frac{l^2+1}{2}m+\frac{l+1}{2})n+\frac{l^2+1}{2}m+\frac{l+1}{2}=s(q+1)\leq
x+yq \leq (s+m)(q+1)<(\frac{l^2+1}{2}m+\frac{l+1}{2}+1)\cdot n$;

$(13m+3)\cdot n<(13m+3)n+13m+3=s(q+1)\leq x+yq \leq
(s+m)(q+1)=(13m+3)n+26m^2+9m+2<(13m+4)\cdot n$; and using this
method, one can obtain that
 if $ x,y\in I_{im}$, then
$(\frac{l^2+1}{2}m+\frac{l+1}{2}+i)\cdot n<(s+im+1)\cdot (q+1)\leq
x+yq \leq (s+(i+1)m)(q+1)<(\frac{l^2+1}{2}m+\frac{l+1}{2}+i+1)\cdot
n$, where $0\leq i\leq \frac{l+1}{2}m+1$ if $l=10t+3$ and $0\leq
i\leq \frac{l+1}{2}m+2$ if $l=10t+7$.

Hence, there is no skew symmetric cyclotomic cosets, and any two
cyclotomic  coset do not form a skew asymmetric pair in $T\setminus
\{C_{s+3m},C_{s+2m}\}$. That implies that
$T_{ss}=\{C_{s+\frac{l+3}{4}m},C_{s+\lfloor \frac{l}{10} \rfloor
m}\}$ if  $l=10t+3$ and $|T_{ss}|=4$;
$T_{ss}=\{C_{s+\frac{(\frac{l-1}{2}+\lceil\frac{l}{10}\rceil)}{2}m},C_{s-\lceil\frac{l}{10}\rceil
m-1}\}$ if  $l=10t+7$  and $|T_{ss}|=4$ for $0 \leq k \leq 7m$, when
the defining set $T=\bigcup_{i=0}^{k}C_{s+i}$, where $0\leq k\leq
\frac{l+1}{2}m+1$ if $l=10t+3$; $0\leq k\leq \frac{l+1}{2}m+2$ if
$l=10t+7$.\\

 {\bf  Theorem 3.4:}   Let  $q$ be an odd prime power in the form of
$q=am+l$,  $a=\frac{(l^2+1)}{5}$,  $l=10t+3$ or $l=10t+7$ is an odd
number, where $m,t$ is a positive integer. Then there exists a q-ary
$[[\frac{q^{2}+1}{a},\frac{q^{2}+1}{a}-2d+6,d;4]]$- EA-quantum MDS
codes, where $(\frac{l-1}{2}+\lceil\frac{l}{10}\rceil)m+5\leq d \leq
(l+1)m+5$ is  odd.

{\bf  Proof:} Consider the cyclic codes over $F_{q^{2}}$ of length
$n=\frac{q^2+1}{a}$ with defining set Let
$T=\bigcup_{i=0}^{k}C_{s+i}$, where $0\leq k\leq \frac{l+1}{2}m+1$
if $l=10t+3$; $0\leq k\leq \frac{l+1}{2}m+2$ if $l=10t+7$. Since $q$
is an odd prime power  in the form of  $q=am+l$,
$a=\frac{(l^2+1)}{5}$,  $l=10t+3$ or $l=10t+7$ is an odd number,
where $m,t$ is a positive integer, by Lemma 3.3, there is
$c=|T_{ss}|=4$ if $\frac{l+1}{4}m\leq k\leq \frac{l+1}{2}m+1$ if
$l=10t+3$; $\frac{l+1}{4}m\leq k\leq \frac{l+1}{2}m+2$ if $l=10t+7$.
Since every $q^{2}$-cyclotomic coset $C_{s+x}=\{s+x,s-x-1\}$, $0\leq
x\leq s-1$ and $s=\frac{n+1}{2}$, we can obtain that $T$ consists of
$2(k+1)$ integers $\{s-k-1,\cdots,s-2,s-1,s,s+1,\cdots,s+k\}$. It
implies that $\mathcal{C}$ has minimum distance at least $2k+1$.
Hence, $\mathcal{C}$ is a $q^{2}$-ary cyclic code with parameters
$[n,n-2(k+1),\geq 2k+3]$. Combining Theorem 2.6 with EA-quantum
Singleton bound, we can obtain a EA-quantum MDS code with parameters
$[[\frac{q^{2}+1}{a},\frac{q^{2}+1}{a}-2d+6,d;4]]_{q}$, where
$(\frac{l-1}{2}+\lceil\frac{l}{10}\rceil)m+5\leq d \leq (l+1)m+5$ is
odd.

\begin{center}
Table 2 EAQMDS codes with $n=\frac{q^{2}+1}{a}$, $a=\frac{l^2+1}{5}$ \\
\begin{tabular}{lllllllllll}
  \hline
          &q           &$[[n,k,d;4]]_{q}$                  &d     is odd                           \\
\hline
 $q=m+l$,                 & 17                     &$[[29,35-2d,d;4]]_{17}$                &$9\leq d \leq 13$         \\
$a=\frac{l^2+1}{5}$,
 $l=7$                    & 27            &$[[73,79-2d,d;4]]_{27}$                &$13\leq d \leq 21$             \\

                        &   37            &$[[137,142-2d,d;4]]_{37}$                &$17\leq d \leq 29$         \\

  \quad$l=13 $            &   47              &$[[65,71-2d,d;4]]_{47}$        &$13\leq d \leq 19$              \\

            &    81                   &$[[193,199-2d,d;4]]_{81}$         &$21\leq d \leq 33$               \\

            &   149                    &$[[653,659-2d,d;4]]_{149}$       &$37\leq d \leq 61$             \\

\quad $l=17$              &   191                   &$[[629,635-2d,d;4]]_{191}$              &$35\leq d \leq 59$              \\

            &    307                   &$[[1625,1631-2d,d;4]]_{307}$               &$55\leq d \leq 95$               \\

 \quad $l=27$          &   173                   &$[[205,211-2d,d;4]]_{173}$              &$21\leq d \leq 33$              \\

 \hline
  \end{tabular}
\end{center}

\section{SUMMARY}
\label{sec:1}

In this paper, by analysing the concept of decomposition of the
defining set of cyclic codes, we construct two families of $q$-ary
entanglement-assisted quantum MDS codes
$[[\frac{q^{2}+1}{a},\frac{q^{2}+1}{a}-2(d-1)+4,d;4]]$, where q is a
prime power  in the form of  $q=am+l$ based on classical cyclic MDS
codes by exploiting 4 pre-shared maximally entangled states. $a$,
 $l$ and $m$ is in the following:

 (i) $a=l^2+1$,  $l$ is a odd number and $m$ is integer number;

(ii) $a=\frac{l^2+1}{5}$,  $l=10m+3$ or $l=10m+7$ is a odd number
and $m$ is integer number.

In Table 3, we list the $q$-ary entanglement-assisted quantum MDS
codes constructed in this paper. For $l=3$, consumed four pre-shared
maximally entangled states, each EA-quantum MDS code of length
$n=\frac{q^2+1}{10}$  has twice larger minimal distance than  the
standard QMDS code of code length constructed in Ref.\cite{Chen}.
Comparing the parameters with all known $q$-ary EA-quantum MDS
codes, we find that all these constructed EA-quantum MDS  codes  are
new in the sense that their parameters are not covered by the codes
available in the literature.

\begin{center}
Table 3 New parameters of EAQMDS codes \\
\begin{tabular}{lllllllllllllll}
  \hline
  $a$               & q                &$[[n,k,d;c]]_{q}$                                                      &Distance (d is odd) \\
\hline
  $l^{2}+1$         &$q=am+l$        &$[[\frac{q^{2}+1}{a},\frac{q^{2}+1}{a}-2d+6,d;4]]$               &$(l+1)m+3\leq d \leq (3l-4)m+3$  \\
                    &                &                                                                    &   \\

  $\frac{l^{2}+1}{5}$  &$q=am+l$      &$[[\frac{q^{2}+1}{a},\frac{q^{2}+1}{a}-2d+6,d;4]]$     &$(\frac{l-1}{2}+\lceil\frac{l}{10}\rceil)m+5\leq d \leq (l+1)m+5$  \\
                       & $l=10m+3$ or     &                                                         &        \\
                       & $l=10m+7$     &
                       &\\

  \hline
  \end{tabular}
\end{center}

\section{Acknowledgment}
This work is supported by the National Natural Science Foundation of
China under Grant No.11801564 and No.61373171, the National Key R\&D
Program of China under Grant No. 2017YFB0802400, 111 Project under
grant No.B08038.

\bibliography{wenxian}
\end{document}